\documentclass[12pt,preprint]{aastex}

\newcommand{\HI}{\mbox{H\,{\sc i}}}
\newcommand{\kms}{\mbox{km~s$^{-1}$}}

\slugcomment{To Appear in ApJ}

\shorttitle{}
\shortauthors{E. Iodice et al.}

\begin{document}


\title{Stellar kinematics for the central spheroid in the 
Polar Disk Galaxy NGC~4650A \footnote{
Based on data collected with the FORS2 spectrograph, mounted at the UT4 of the
Very Large Telescope at Cerro Paranal, Chile, operated by ESO, during
observing run 70.B-0277}}


\author{E. Iodice\altaffilmark{1}, M. Arnaboldi\altaffilmark{2,3}, R. P. Saglia\altaffilmark{4}, L.S. Sparke\altaffilmark{5}, O. Gerhard\altaffilmark{4,6}, J.S. Gallagher\altaffilmark{5}, F. Combes\altaffilmark{7}, F. Bournaud\altaffilmark{7}, M. Capaccioli\altaffilmark{8,9}, K.C. Freeman\altaffilmark{10}}
 
\altaffiltext{1}{INAF-Oss. Astr. di Capodimonte, via Moiariello 16, 80131 Napoli, Italy (iodice@na.astro.it)}
\altaffiltext{2}{INAF-Oss. Astr. di Torino, Strada Osservatorio 20, Pino Torinese, Italy (arnaboldi@to.astro.it)}
\altaffiltext{3}{European Southern Observatory, Karl-Schwarzschild-Strasse 2, D-85748 Garching bei Munchen, Germany}
\altaffiltext{4}{Max-Planck Institut fuer extraterrestrische Physik, Giessenbachstrasse, 85748 Garching, Germany (saglia@mpe.mpg.de)}
\altaffiltext{5}{Univ. of Wisconsin, Department of Astronomy, 475 N. Charter St., Madison, WI, USA}
\altaffiltext{6}{Astronomisches Institut der Universitat Basel, Venusstrasse 7, CH-4102 Binningen, Switzerland}
\altaffiltext{7}{Observatoire de Paris, LERMA, 61 Av. de l'Observatoire, 75014, Paris, France}
\altaffiltext{8}{Dip.Scienze Fisiche Univ. di Napoli Federico II, via Cintia ed. G, 80126 Napoli, Italy }
\altaffiltext{9}{VST Center at Naples (VSTceN) at INAF-Oss. Astr. di Capodimonte, Via Moiariello 16, 80131 Napoli, Italy }
\altaffiltext{10}{RSAA, Mt. Stromlo Observ., Cotter Road, Weston Creek, ACT 2611, Australia}


\begin{abstract}
We have obtained high angular resolution, high signal-to-noise spectra
of the Calcium triplet absorption lines on the photometric axes of the
stellar spheroid in the polar disk galaxy NGC~4650A.  Along the major
axis, the observed rotation and velocity dispersion measurements show
the presence of a kinematically decoupled nucleus, and a flat velocity
dispersion profile. The minor axis kinematics is determined for the
first time: along this direction some rotation is measured, and the
velocity dispersion is nearly constant and slightly increases at
larger distances from the center. The new high resolution kinematic
data suggest that the stellar component in NGC~4650A resembles a
nearly-exponential oblate spheroid supported by rotation.  The main
implications of these results on the previous mass models for
NGC~4650A are discussed. Moreover, the new kinematic data set
constraints on current models for the formation scenarios of Polar
Ring Galaxies (PRGs), supporting a slow accretion rather then a
secondary strong dissipative event.

\end{abstract}


\section{Introduction}
\begin{figure}
\includegraphics[width=9cm]{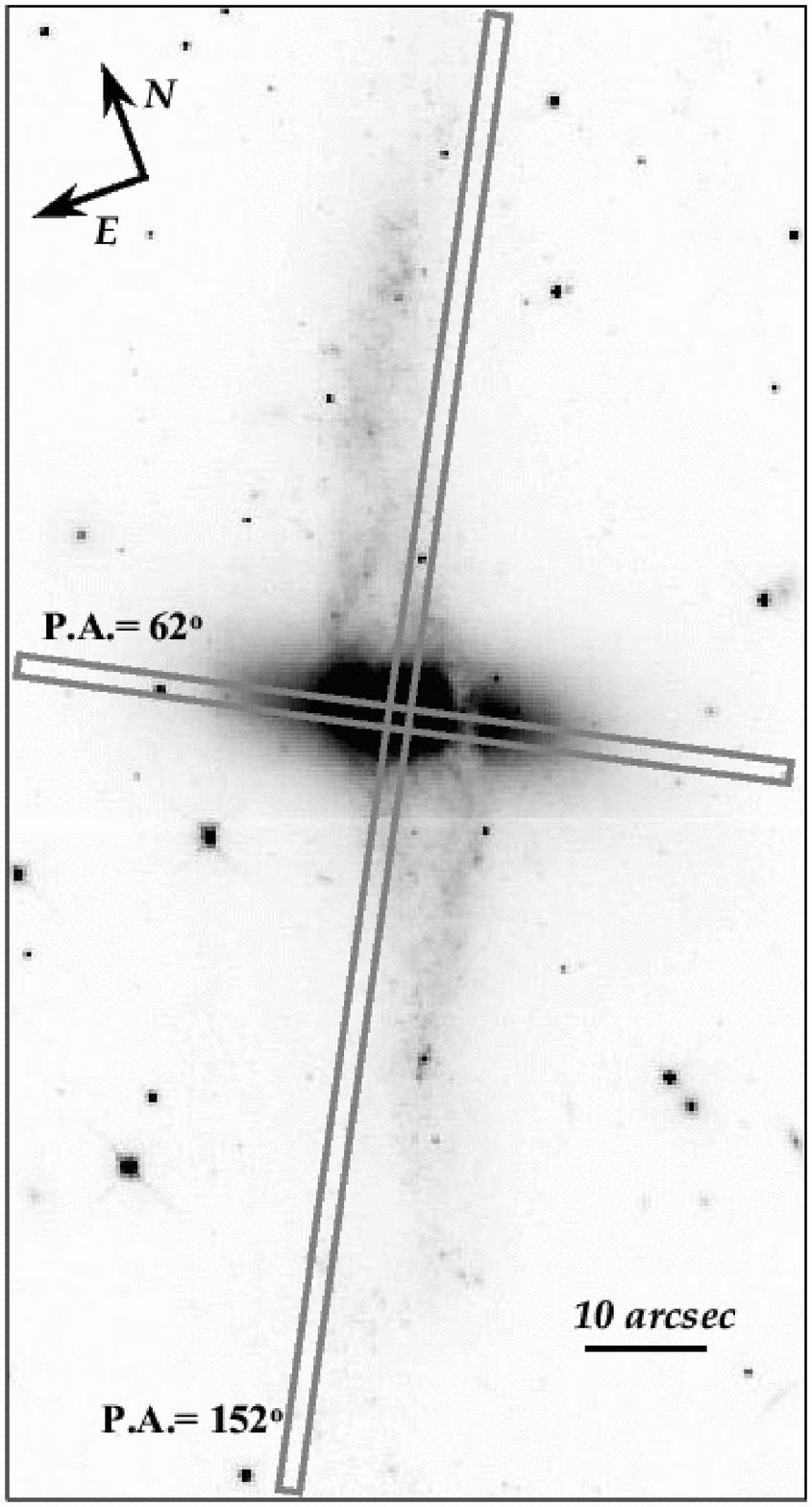}
\caption{HST I band image of NGC~4650A, with the slit 
positions overlaid. 
\label{fig1} }
\end{figure}

Studies of galaxy formation and evolution both in the Local Group and
in the high-redshift universe (Conselice et al. 2003) provide
increasing evidence that mergers play a role in the formation of
early-type galaxies, in clusters and in the field (Schweizer 1999,
Hibbard \& Yun 1999). The merging of two disk galaxies produces a
spheroidal remnant with physical properties, such as density profiles,
mean velocity dispersion and surface brightness, similar to those
observed for elliptical and early-type disk galaxies (Toomre \& Toomre
1972; Gerhard 1981; Barnes \& Hernquist 1992; Bekki 1998a; Naab \&
Burkert 2003; Bournaud et al. 2005). These kind of mergers may also
produce a PRG (Bekki 1998b; Bournaud \& Combes
2003), like NGC~4650A (Fig.\ref{fig1}).

PRGs generally contain a central featureless stellar spheroid and an
elongated structure, the ``polar ring'', which in projection appears
close to the inner spheroid's meridian plane. In all PRGs, the \HI\
gas is associated with the polar structure and not with central
stellar spheroid; furthermore the stars in the central spheroid and
the gas and stars in the polar structure rotate in two nearly
perpendicular planes. The studies of PRGs promise to yield detailed
information about some of the processes at work during galaxy merging
and on the shape of dark matter halos in galaxies (Schweizer, Whitmore
\& Rubin 1983, Sackett \& Sparke 1990, Sackett et al. 1994, Combes \&
Arnaboldi 1996, Iodice et al. 2003).

NGC~4650A is the prototype for PRGs. Its luminous components, inner
spheroid and polar structure, have been studied with optical and
near-infrared (NIR) photometry, optical spectroscopy, and in the radio
(\HI\ 21 cm line and continuum).  The surface brightness profile of
the central spheroid is described by an exponential law, with a small
exponential nucleus; its integrated optical vs. NIR colors are similar
to those of an intermediate age stellar population (Iodice et
al. 2002; Gallagher et al. 2002). Previous absorption line spectroscopy
at optical wavelengths showed a substantial rotation along the major
axis, with $v_{max} \simeq 100$ \kms (Whitmore et al. 1987, Sackett et
al. 1994), while the velocity dispersion measurements were plagued by
systematic errors, caused by both low angular and low spectral
resolution. Best estimates gave $\sigma_0 \simeq 60$ \kms.

The polar structure in NGC~4650A has been shown to be a disk, rather
than a ring. Its stars and dust can be reliably traced inward within
the stellar spheroid to $\sim 1.2$ kpc radius from the galaxy nucleus
(Iodice et al. 2002; Gallagher et al. 2002). Emission and absorption
line optical spectra show that it is an extended stellar disk rather
than a narrow ring (Swaters \& Rubin 2003): both rotation curves show
a linear inner gradient and a plateau, as expected for a disk in
differential rotation. Furthermore the \HI\ 21 cm observations
(Arnaboldi et al. 1997) show that the gas is 5 times more extended
than the luminous polar structure, with a position-velocity diagram
very similar to those observed for edge-on gaseous disks.

The question about the shape of the dark halo of NGC~4650A is still
open. Because of the large error-bars in the velocity dispersion
profile, dynamical models by Sackett et al. (1994) and Combes \&
Arnaboldi (1996), which differ in the orientation of the main axes and
flattening of the dark halo, were both compatible with the observed
data. The polar disk luminous mass distribution along the meridian
plane may also affect the inner stellar kinematics and induce a
triaxial-like perturbation to the axisymmetric stellar mass
distribution. Kinematic information along independent position angles
on the sky may therefore be required for a reliable mass
model. Furthermore, moving to longer wavelength may help reducing the
contamination from the dust (Arnaboldi et al. 1995).

An 8 meter class telescope, with a spectrograph like FORS2 and a high
efficiency holographic grism in the 1 micron wavelength range shall
allow us to study the stellar motions of the NGC~4650A inner spheroid,
via the absorption line spectroscopy of the Calcium triplet (CaT)
lines, at 8498 \AA, 8542 \AA, 8662 \AA.  These lines are the strongest
features in the stellar continuum for a large variety of stellar types
(Cenarro et al. 2001).  In this work, we shall present our
measurements for the radial velocity, the velocity dispersion profile
and the Gaussian Hermite parameters $H_3,\, H_4$ along the main
photometric axes of the central spheroid in NGC~4650A.  We then
discuss their implication on the proposed mass models and formation
scenario for PRGs. In what follow, we shall adopt 2905 \kms as the
heliocentric systemic velocity of NGC~4650A (Arnaboldi et
al. 1997) and H$_0 = 70$\kms Mpc$^{-1}$, which implies $1''= 201$ pc.


\section{Observations and data reduction}\label{data}

The spectra were obtained with FORS2@UT4, on the ESO VLT, in service
mode, during the observing run 70.B-0277. FORS2 was equipped with the
MIT CCD~910, with an angular resolution of $0''.125$ pixel$^{-1}$ and
a binning of two pixels, so the final scale is $0''.25$
pixel$^{-1}$. Spectroscopic observations were carried out with a slit
$1''.6$ wide and $6'.8$ long, and grism GRIS-1028z+29 in the
$7730-9480$ \AA\ wavelength range, which covers the red-shifted
absorption lines of the CaT at the systemic velocity of NGC~4650A. The
nominal spectral resolution is 0.86 \AA pixel$^{-1}$.  Spectra were
acquired at the position angles of the photometric major
(PA=$62^{\circ}$) and minor (PA=$152^{\circ}$) axes (Iodice et
al. 2002) of the stellar spheroid (Fig.\ref{fig1}). A total
integration time of 5 hours was acquired for the major axis and 4.5
hours for the minor axis, with an average seeing of $0''.9$ and of
$0''.83$ respectively. In the final median averaged spectra we reach a
limiting surface brightness of $\mu_I = 23$ mag arcsec$^{-2}$ at $r =
20''.5$ along the spheroid major axis. Spectra of standard template
stars of the K0III and K3III spectral types, were also acquired and
trailed along the slit.

The frames were bias-subtracted, flat-fielded and calibrated using
standard IRAF tasks, with Helium, Argon and Neon arc-lamps taken for
each observing night. The final solution for the wavelength
calibration reaches an rms of 0.07 \AA. The instrumental resolution
measured from the arclamp spectra is $\sigma =2.1$ \AA; at the CaT
wavelengths, this is equivalent to a velocity resolution of $\sigma =
70$ \kms.

\begin{figure}
\includegraphics[width=13cm]{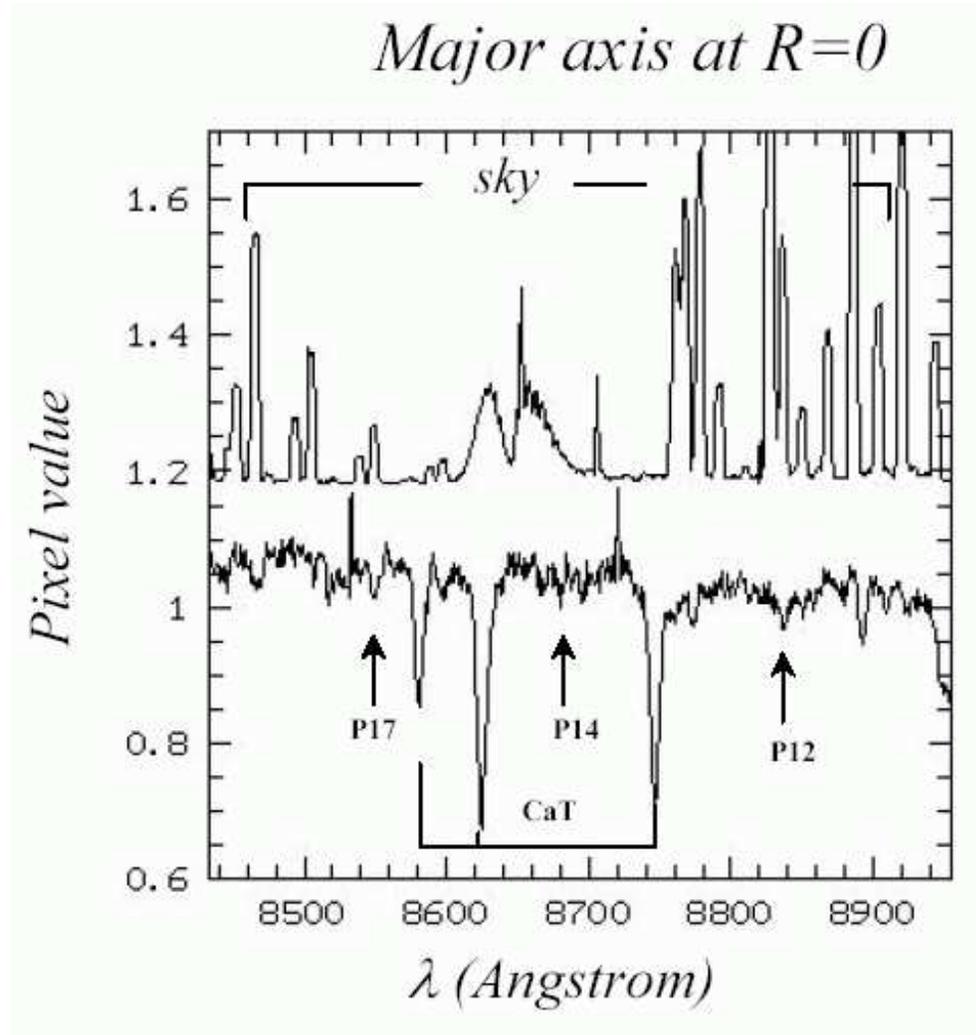}
\caption{Normalized spectra along the major axis in the center of
spheroid (on the bottom), where CaT and PaT lines are marked, and
along the sky. Note that the emission lines in the sky are stronger
for $\lambda \ge 8740 \dot{A}$, where the Paschen line P12 is located.
\label{spectra}}
\end{figure}

The sky spectrum was extracted at the outer edges of the slit, where
there is no galaxy light, and subtracted off each row of the two
dimensional spectra by using the IRAF task BACKGROUND in the
TWODSPEC.LONGSLIT package. On average, a sky subtraction better than
$1\%$ was achieved; in the region for $\lambda \ge 8740$ \AA, where
the sky emission lines are stronger and blended (Fig.~\ref{spectra}),
residuals are less than the $2\%$.  After sky subtraction, for each
PA, the scientific frames were co-added to a final median averaged 2D
spectrum.  The final spectrum from the major axis slit at the center
of NGC~4650A is shown in Fig.~\ref{spectra}.

The final steps of the data-processing were {\it i)} the binning of
the spectra along the spatial direction to achieve a minimum
signal-to-noise (S/N) $\ge 50$ at all radii (which is the S/N measured
at the last data points, while the central pixels have higher S/N by
up to a factor of 4) and to leave no more than 2 independent data
points within the seeing disk; {\it ii)} removal of the galaxy
continuum by fitting a forth order polynomial (for a detailed
description of the procedures see Bender, Saglia \& Gerhard 1994).

\section{New kinematics of the central spheroid: systematics and
error estimate}\label{kin}

The {\it Line-of-Sight Velocity Distribution} (LOSVD) was recovered
from the continuum-removed spectra using the Fourier Correlation
Quotient (FCQ) method (Bender 1990; Bender, Saglia \& Gerhard 1994).
By assuming that the LOSVD is described by a Gaussian plus third- and
fourth-order Gauss-Hermite functions (van der Marel \& Franx 1993;
Gerhard 1993), the rotational velocities $v$, velocity
dispersions $\sigma$ and first order deviations from Gaussian
profiles, $H_3$ and $H_4$, were derived at each radius.  The best fit
to the spheroid spectrum was obtained with the K3III template,
while the K0III stellar template was discarded because of a
significant mismatch.
  
Both statistical and systematic errors for our radial velocity,
$\sigma $ and the $H_3$ and $H_4$ parameters were studied via Monte
Carlo simulations (Bender, Saglia \& Gerhard 1994; Mehlert et
al. 2000).  It turns out that, for all these kinematic quantities, the
statistical errors dominate with respect to systematic errors at all
radii.

\subsection{Influence of Paschen lines}\label{Paschen}

The three lines of the Paschen sequence, P13, P15 and P16 at
$\lambda_{rest}=8665$ \AA, 8545 \AA\ and 8502 \AA, lie within the red
wings of the CaT lines (Cenarro et al. 2001), and they may affect the
measured $v$ and $\sigma$, if the adopted stellar template does not
match them. Therefore we need to quantify the equivalent width of the
Paschen hydrogen lines, and their strength along the slit, and compare
them with the features in the stellar template spectrum, before we
derive the spheroid stellar kinematics in NGC~4650A.

We measured the equivalent width (EW) of the unblended Paschen lines
P14, P17, hereafter Pa(17-14), at $\lambda_{rest}= 8598$ \AA, 8467
\AA, and the CaII $\lambda_{rest}= 8498$ \AA, 8542 \AA\ along the slit
for the two axes; the EW of the Paschen absorption line P12
($\lambda_{rest}= 8750$ \AA) cannot be measured reliably because of
some strong sky lines residuals which affect the estimates of the
continuum flux (Fig.\ref{spectra}).

\begin{figure}
\includegraphics[width=17cm]{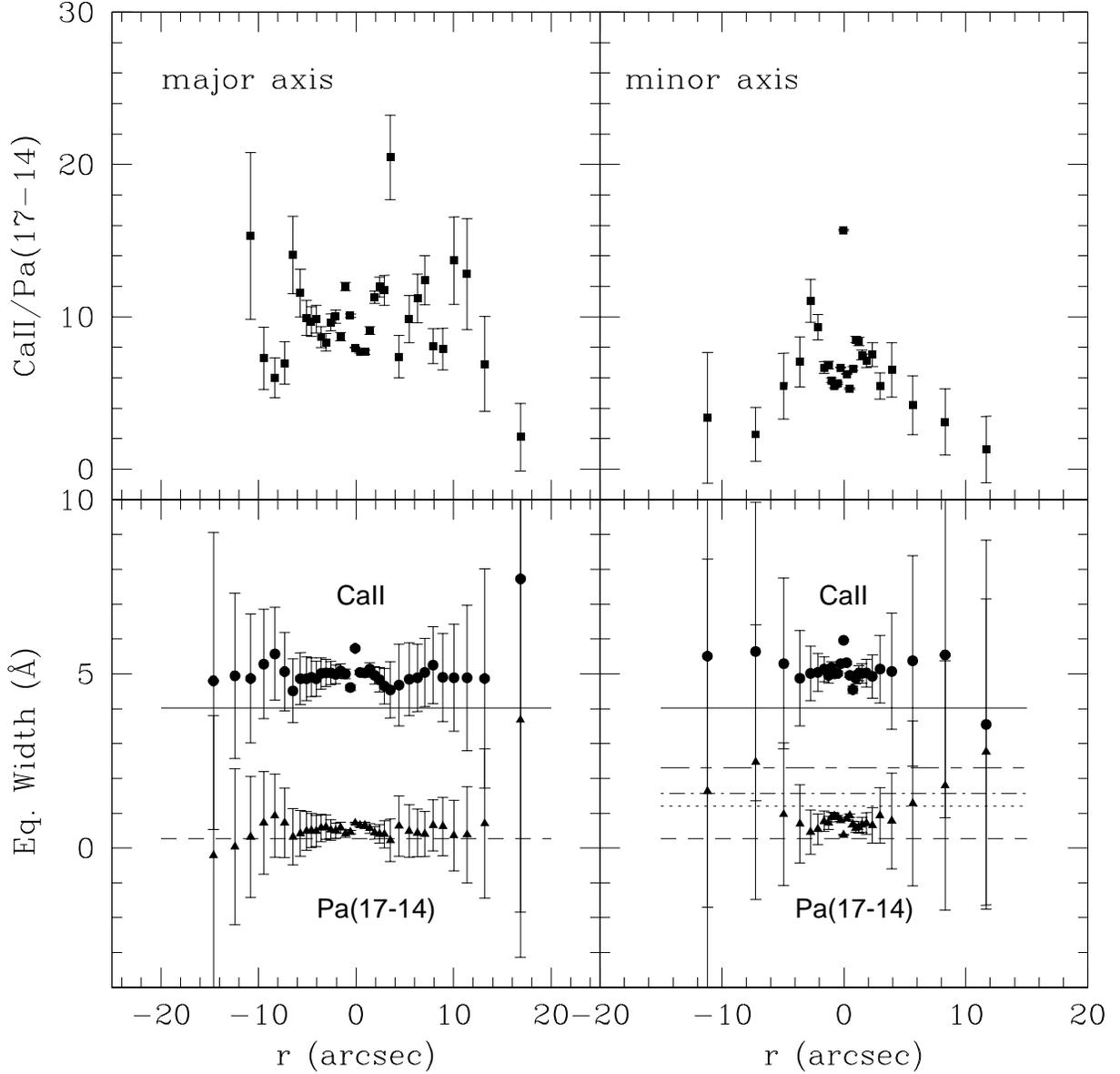}
\caption{ Top panels - Equivalent width (EW) ratio between the CaII and
Pa(17-14) lines versus distance from the galaxy center along the major
(left) and minor (right) axis. Bottom panels - EW of the
CaII lines (circles) and Pa(17-14) lines (triangles) along the major
(left) and minor (right) axis. Solid and short-dashed lines indicate
the CaII and Pa(17-14) EW respectively in the synthetic galaxy
spectrum made by a pure K star template. In bottom-right panel, the
Pa(17-14) EWs for different synthetic spectra are also
shown. Synthetic spectra were obtained by combining stars of spectral
type A and K (see also Sec.\ref{Paschen}): dotted line is for $20\%$ A
+ $80\%$ K, dashed-dotted line is for $30\%$ A + $70\%$ K, short dash
- long dash is for $50\%$ A + $50\%$ K .
\label{PaT}}
\end{figure}

In Fig.\ref{PaT}, we show the Pa(17-14), the CaII EWs
and their ratios along the two axes. In the central region of the
NGC~4650A spheroid and along its major axis, the Pa(17-14) and the
CaII EWs are similar to those computed for a synthetic
galaxy spectrum using the K3III template star. They
are consistent with the values of EWs observed for the stellar integrated
light of early-type galaxies (Saglia et al. 2002; Falcon-Barroso et
al. 2003).

Along the minor axis, at distances larger than $5''$, the PA(17-14)
vs.  CaII EW ratio seems to reach values which are comparable to those
observed for star forming galaxies (Saglia et al. 2002). The larger
values for the Pa(17-14) EW at these radii may indicate a contribution
to the light from young stars in the polar disk.  

In order to test whether the contribution of a young stellar
population may affect the kinematics, we produced synthetic spectra
combining stars of spectral type A and K, and measured the the
PA(17-14) vs. CaII EW ratio. Fig.\ref{PaT} shows that a template with
an increasing contribution from an A-type spectrum, from a $20\% A +
80\% K$ to a $50\% A + 50\% K$, has Pa(17-14) EWs which are matching
the larger values measured in the outer regions of the spheroid minor
axis. Once this synthetic spectrum is broadened to 80 \kms dispersion
and analyzed using a pure K template with the FCQ method, the measured
velocity dispersion is then 86 \kms, 90 \kms and 100 \kms for the
$20\%A+80\%K$, $30\%A+70\%K$ and $50\%A+50\%K$ cases.  

Along the spheroid minor axis at $4''<r<6''$ and $r>6''$ the Paschen
lines strength are consistent with those of $20\%A+80\%K$ and
$50\%A+50\%K$ templates; from our tests, the effects caused by the
contribution from younger stars may cause the measured $\sigma$ to be
overestimated of about 6\kms and 20\kms respectively.

From the present analysis we can conclude that the K3III template is
adequate for the kinematic analysis in the center and along the major
axis. In the outer regions of the minor axis we found that the young
stars in the polar disk might cause an increase in the Paschen lines
strength (see also Sect.\ref{min}). Nonetheless, given the large
errorbars of the Pa(17-14) EWs, a pure K template is still compatible
with the data and it is adopted to measure the kinematics for both
axes and at all radii.

\subsection{Instrumental resolution}

The kinematics along the major axis of the stellar spheroid in
NGC~4650A were previously measured from absorption line spectra at
optical wavelengths (4100 - 4500 \AA\ and 5085 - 5980 \AA). The latest
measurements based on the Mg {\it b} and the Na {\sc I} absorption
lines were performed by Sackett et al. (1994): they reported a central
value for the velocity dispersion $\sigma_0 \simeq 60$ \kms. This is
similar to our instrumental spectral resolution, as discussed in
Sec.~\ref{data}, and therefore we must evaluate any possible systematic
effects on our measurements.

To this aim, we simulated a synthetic galaxy spectrum using the
stellar template spectrum, with different values of the velocity
dispersion, from 10 to 200 \kms, and with S/N in the range 10 to
120. We then recovered the LOSVD via the FCQ method, using the same
template adopted for our analysis. We found that for S/N$ \ge 50$ and
an input velocity dispersion larger than 50 \kms, the systematic error
is less than 10 \kms; for $\sigma < 50$ \kms, the measured values tend
to the asymptotic value of 40 \kms.  For smaller values of $S/N$, the
systematic errors are larger. Based on these results, we have adopted
a variable spatial binning along the slit which increases with
distance from the galaxy center so that a $S/N \ge 50$ is achieved as
we step to larger distances from the NGC~4650A center. At the galaxy
center, $S/N \sim 200$ .


\section{LOSVD along the principal photometric axes of NGC~4650A}\label{result}

Quantitative photometry of NGC~4650A has shown that the internal
structure of this object is rather complex (Gallagher et al. 2002,
Iodice et al. 2002), and we now wish to examine the correlation
between morphological subcomponents and kinematic signatures. In
Fig.~\ref{fig1} and Fig.~\ref{fig2} we overlay the slit positions on
the HST images of NGC~4650A and in the following section we briefly
describe the luminous subcomponents along both axes, which affect the
stellar light in the spectra. In Fig.~\ref{RCmaj} and Fig.~\ref{RCmin}
we plot the rotation curve, velocity dispersion, the $H_3$ and $H_4$
profiles, for both the major and minor axes of the spheroid in
NGC~4650A.

\subsection{Morphological components along the main axes}\label{phot}

\begin{figure}
\includegraphics[width=15cm]{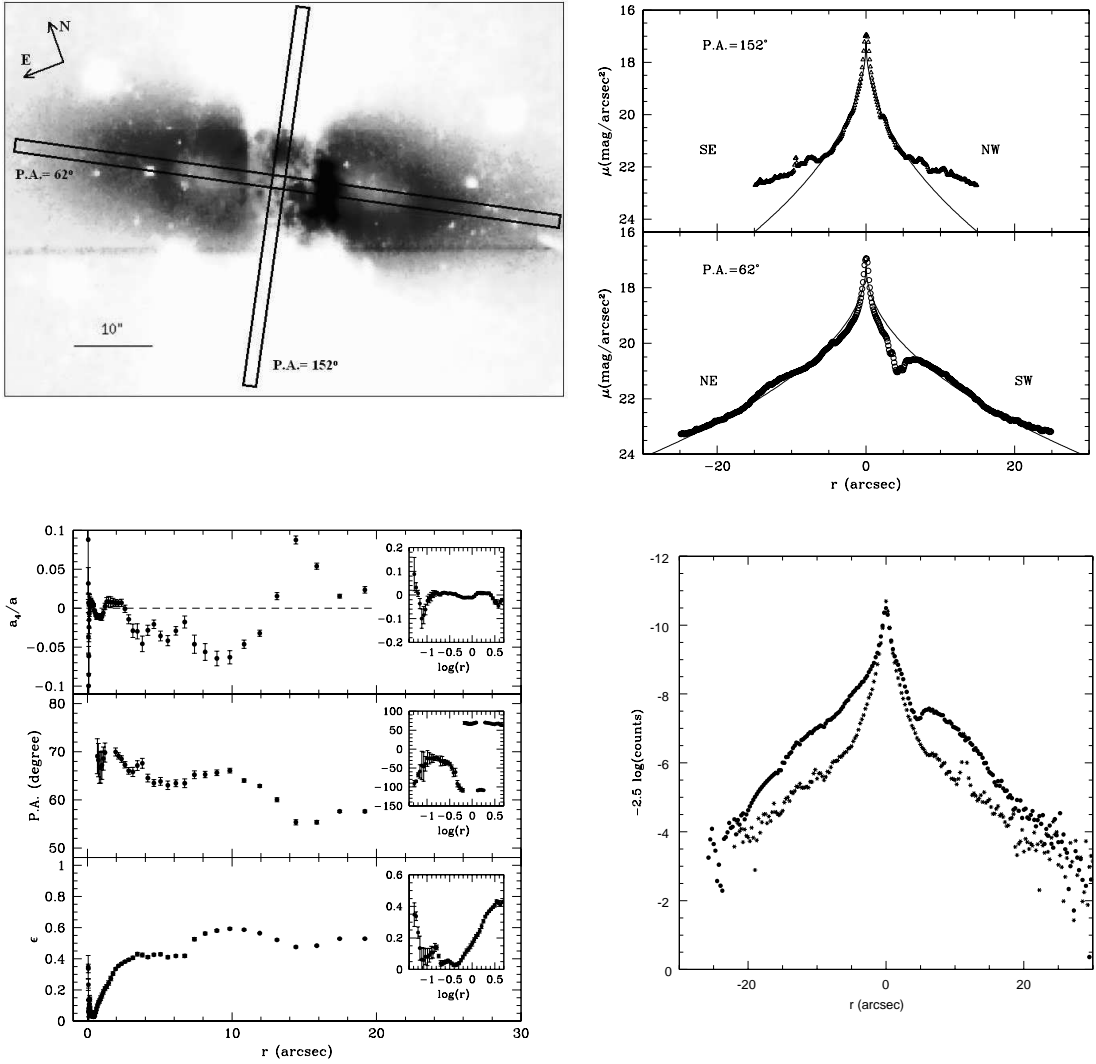}
\caption{Top left panel - Residual image, obtained as the ratio
between the whole galaxy frame and the 2-dimensional model of the
spheroid light distribution in the I band (see Iodice et al. 2002 for
details). Units are intensity; whiter colors correspond to those
regions where the galaxy is brighter than the model, darker colors
corresponds to those regions where the galaxy is fainter than the
model. Top right panel - Surface brightness profiles (open points) and
$r^{1/n}$ fit (solid line) along the major and minor axes of the inner
stellar spheroid (from Iodice et al. 2002); Bottom left panel -
Ellipticity, P.A. profiles and isophotal shape parameter $a_4$ of the
stellar spheroid in I band. Bottom right panel - Uncalibrated
luminosity profiles extracted through the slit along the major
(circles) and minor (asterisks) axes.
\label{fig2} }
\end{figure}

Along the major axis (PA=$62^\circ$), the central spheroid extends to
$r_{outer} = 25''$ from the center in the I band, and $r_{outer}$ is
about $4 \times r_e$; the equivalent effective radius for the central
spheroid in NGC~4650A is $r_e = 6''.7$, and it is evaluated by fitting
the whole I-band light profiles with a Sersic law $I(r) \propto
r^{1/n}$, and $n \sim 2$.  The dust associated with the polar disk,
whose absorption is seen in front of the central spheroid on the SW
side (Fig.\ref{fig1}), affects the light profile along the major
axis at $r \sim 5''$; it is also seen in the extracted light profile
along the slit (Fig.\ref{fig2}, right panel).  For
$r < 5''$, Iodice et al. (2002) showed that the polar disk and the
spheroid light coexist (top-left panel of Fig.\ref{fig2});
within $r<1''$ there is a very luminous, compact nucleus, of about
20 pc ($\sim 0''.1$) in radius and $L_I \sim 8 \times 10^7 L_{\odot}$,
which is superimposed on a more extended and rounder structure, of
about 60 pc $\sim 0''.3$ (Gallagher et al. 2002).

Along the minor axis (PA=$152^\circ$), the light from the polar disk
is present at all radii and at $r > 6''$ from the center becomes
dominant over the spheroid light.  The residual image (by Iodice et
al. 2002) shows that on the SE side of the center the spectrum will
include more light from polar-disk related features than on the NW
side. Within $r < 1''$, the minor axis light is dominated by the
luminous compact nucleus.

\subsection{The major axis kinematics}\label{maj}

\begin{figure*}
\includegraphics[width=11.2cm]{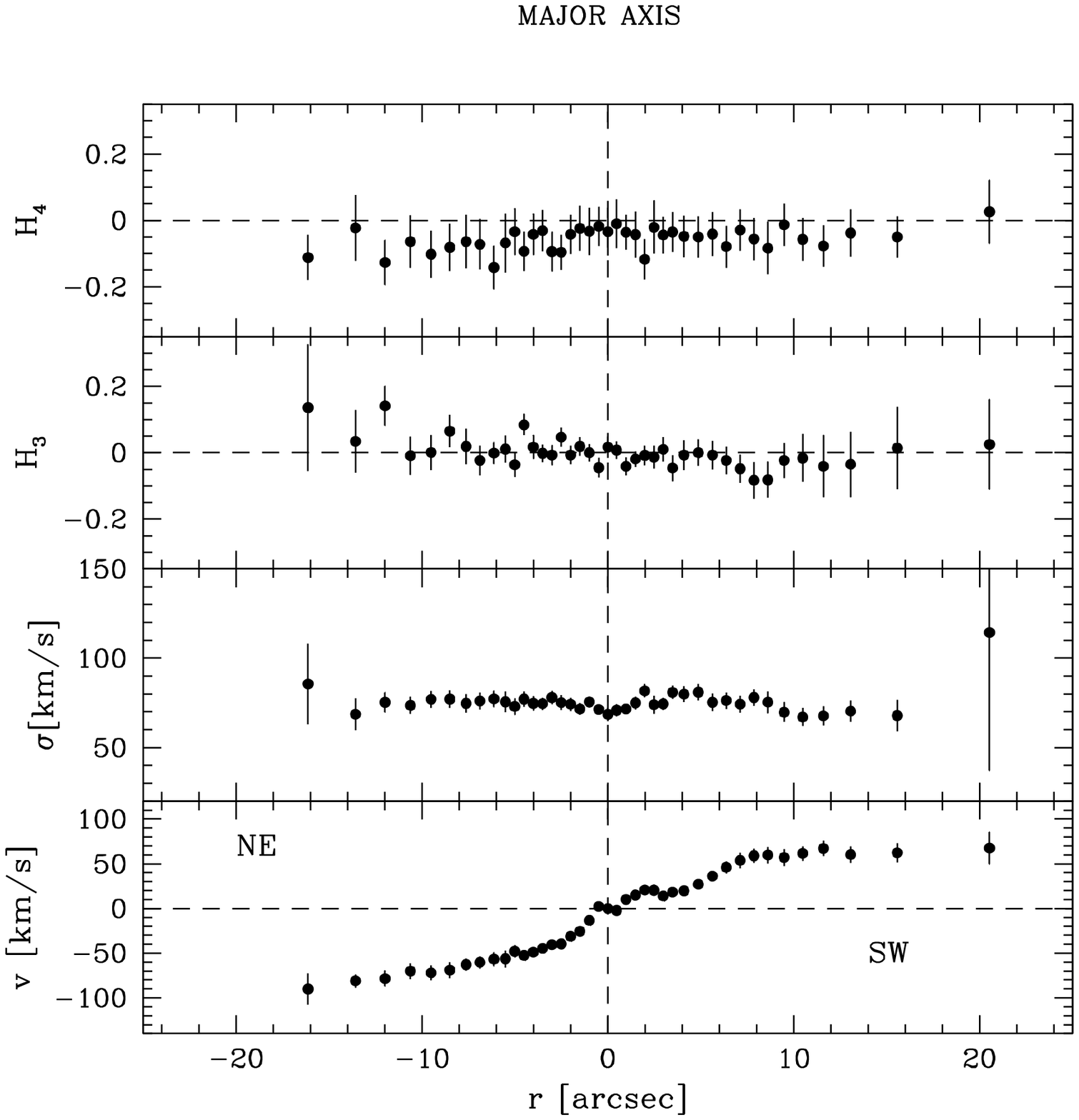}
\includegraphics[width=11.2cm]{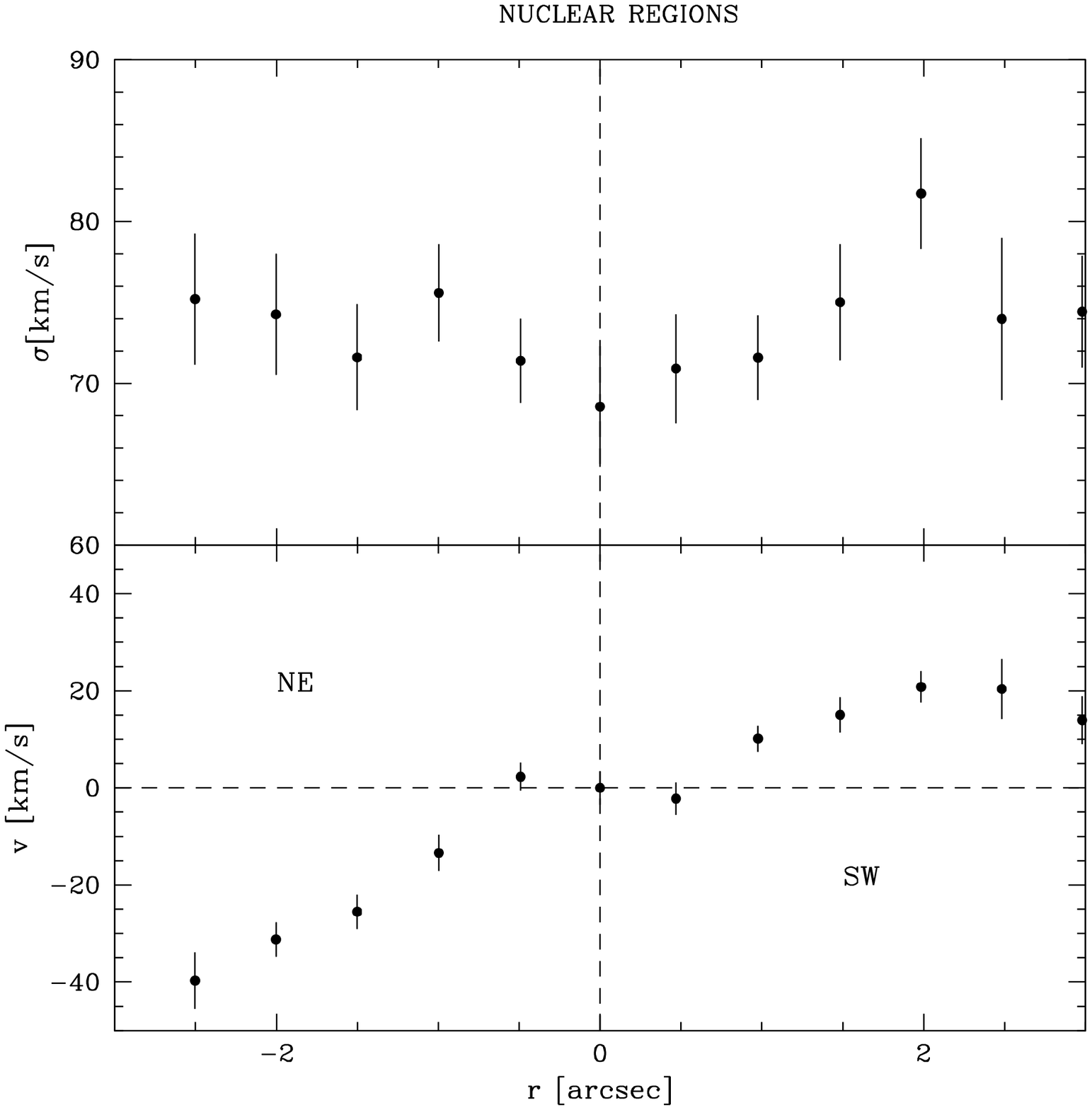}
\caption{Top panel - Rotation curve ($v$), velocity dispersion
($\sigma$), $H_3$ and $H_4$ derived for major axis, $P.A.=
62^{\circ}$, of the inner spheroid in NGC~4650A. Bottom panel - Rotation curve ($v$), velocity
dispersion ($\sigma$) in the nuclear regions of the stellar spheroid.
\label{RCmaj}}
\end{figure*}

The rotation curve is measured out to $20.5''$ ($\sim 4$ kpc),
i.e. about $ 3 \times r_e$ (Fig.~\ref{RCmaj}), which corresponds to a
surface brightness of $\sim 23$ mag arcsec$^{-2}$ in the I band. At
$16''$ ($\sim 3$ kpc) from the center, the velocity increases and it
reaches the maximum value of nearly 100 \kms on the NE side and
slightly lower, $\sim 70$ \kms, on the SW side. The rotation curve is
not symmetric: at $1''< r <5''$ SW (i.e. from 200 pc to 1 kpc), the
rotation curve shows lower velocities than the corresponding NE
side. This was already noticed by Sackett et al. (1994), and a
possible cause is the light contribution from the stars in the polar
disk which passes in front of the spheroid on this side, as we shall
discuss in detail in Sec.~\ref{phot}.

In the nuclear regions (Fig.~\ref{RCmaj} bottom panel), within $0.5''$
radius from the center, the sense of rotation is reversed, and the
measured velocity dispersion decreases to $ \sim 69$ \kms. These
kinematic features correlate with the presence of the bright compact
nucleus and the photometric properties, as for example the variation
of P.A and ellipticity profiles with respect to larger radii
(Fig.\ref{fig2}). Outside the nuclear region, the velocity dispersion
profile is flat, $\sigma \sim 74$ \kms, out to at least $14''$ ($\sim
3$ kpc).

The $H_3$ profile is almost constant, and it is consistent with zero
within the error-bars, from the center out to about $6''$; its value
tends to increase for larger radii. The $H_4$ profile is zero at the
center and then become slightly negative, with an almost constant
value of $-0.05$ on the SW side, and between $-0.05$ and $-0.1$ out to
$16''$ ($\sim 3.2$ kpc) on the NE side. 

\subsection{The minor axis kinematics}\label{min}

\begin{figure*}
\includegraphics[width=11.2cm]{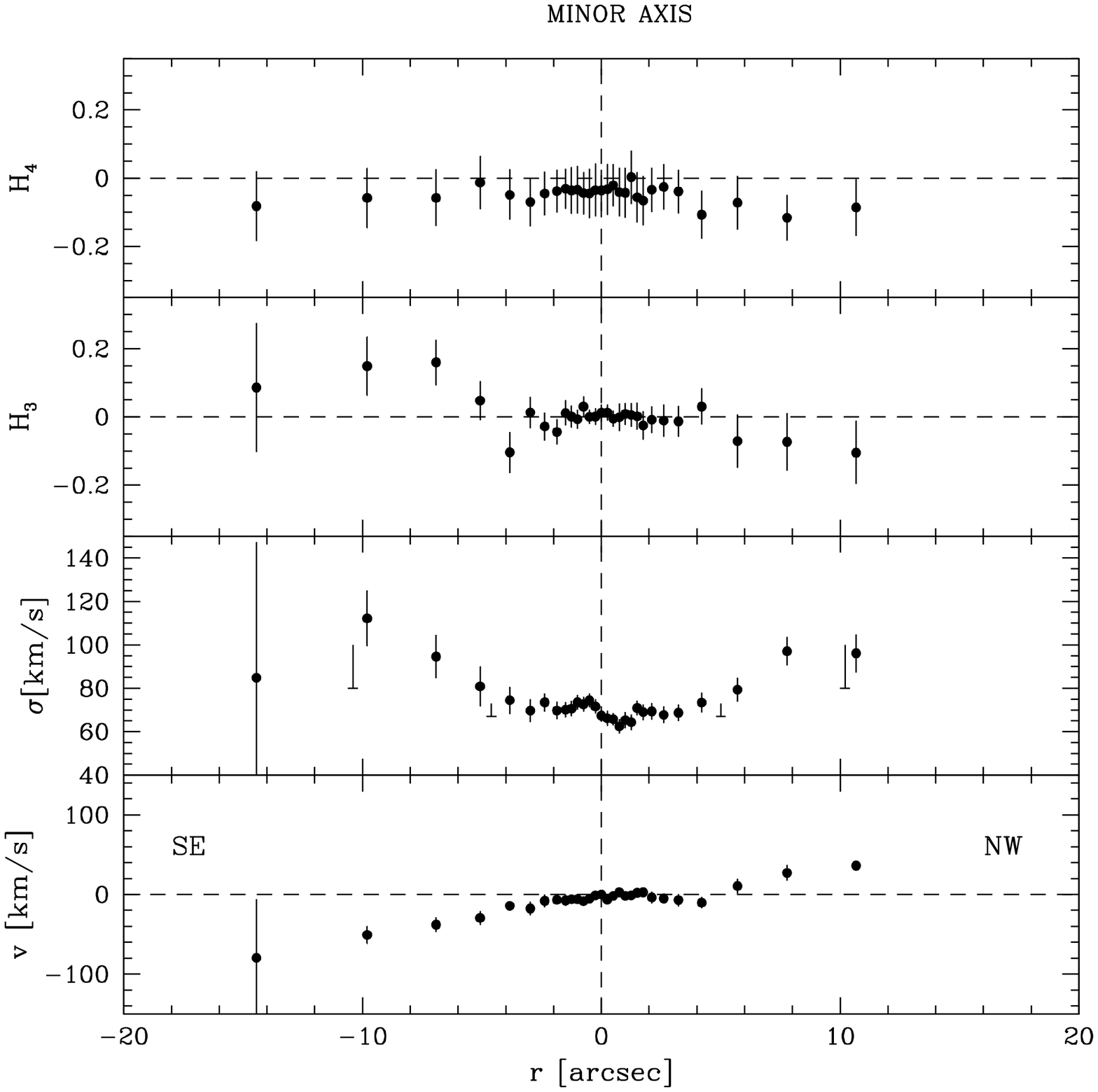}
\includegraphics[width=11.2cm]{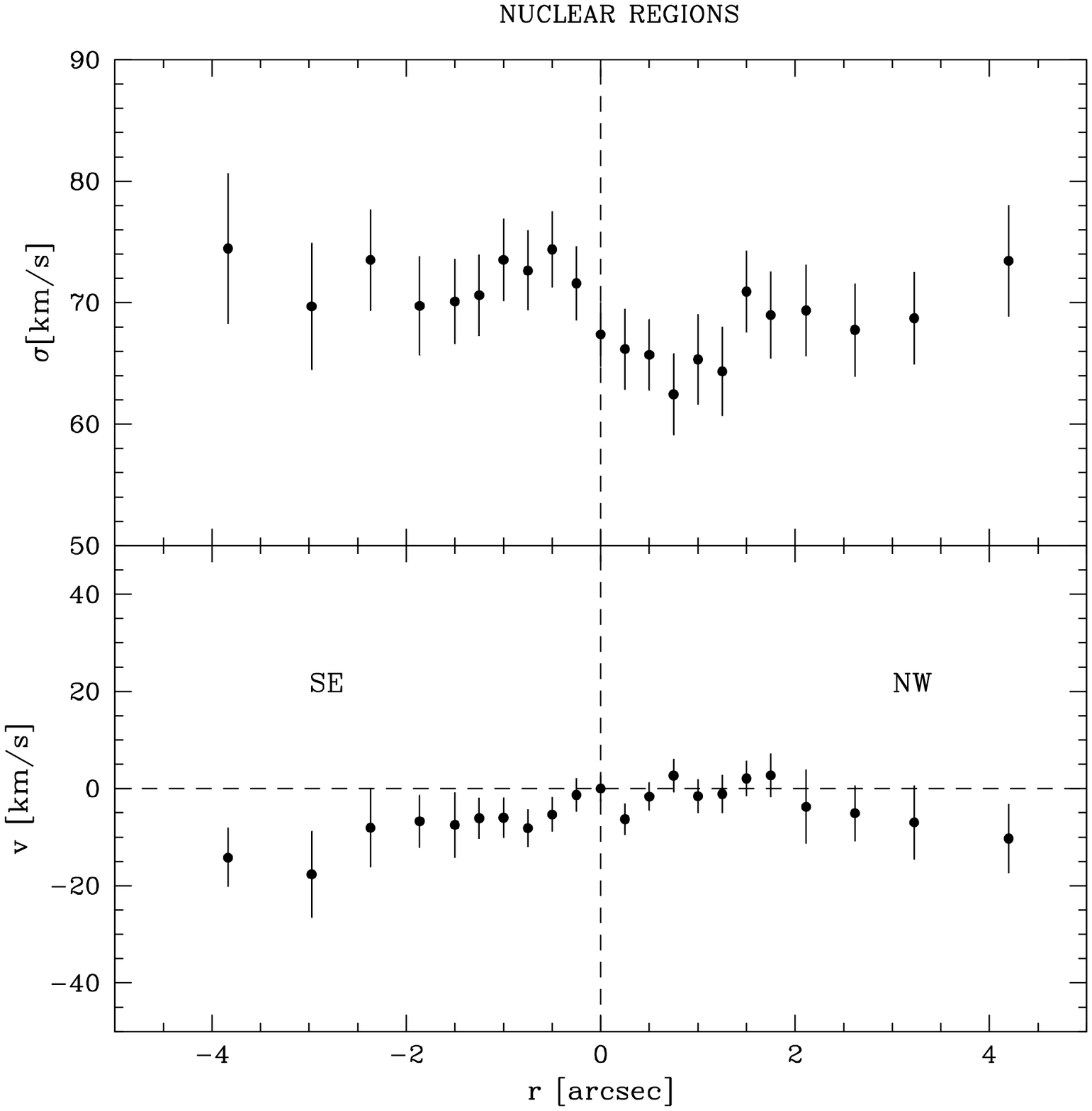}
\caption{Top panel - Rotation curve ($v$), velocity dispersion
($\sigma$), $H_3$ and $H_4$ derived for the minor axis,
$P.A.=152^{\circ}$, of the stellar spheroid. The downwards errorbars
at $r= \pm 5''$ and $r= \pm 10''$ show the systematic effect on the
measured $\sigma$ caused by a young population of stars contributing
up to $20\%$ - $50\%$ of the total light (see Sec.\ref{Paschen} for
details).  Bottom panel - Rotation curve ($v$), velocity dispersion
($\sigma$) in the nuclear regions of the spheroid in NGC~4650A.
\label{RCmin}}
\end{figure*}

On the minor axis, the kinematics can be measured out to $12'' - 15''$
from the center (Fig.~\ref{RCmin} top panel). The central spheroid
dominates the stellar light for $r \le 6'' (\sim r_e)$, and while the
polar disk structure is present at all radii, its surface brightness
dominates the light at $r> 6''$ .

At $r < 4''$, small wiggles with $|v| \simeq 10$ \kms are
visible, but the $v$ profile is generally consistent with no
rotation. At $r > 4''$ the rotation becomes significant and reaches
$v_{max} \simeq 40$ \kms at $10''$, with slightly larger LOS velocities
on the SE.

The measured velocity dispersion is about $\simeq 70$ \kms in the
central parts and then it increases to $\simeq 100$ \kms at $r =
10''$. For $r < 2''$ ($\sim 0.4$ kpc), the velocity dispersion profile
is asymmetric with respect the galaxy center, see the nuclear region
enlargement in Fig.~\ref{RCmin} (bottom panel); larger $\sigma$ values
($ \sim 68 - 74$ \kms) are measured on the SE then on the NW side
($\sigma \sim 62 - 68$
\kms). 

At $r < 4''$ ($\sim 0.8$ kpc), where velocity is nearly zero, the
$H_3$ and $H_4$ profiles are roughly constant, at values equal $0.0$
and $-0.05$ respectively. At $r > 4''$, where $v$ and $\sigma$
increase, the $H_3$ profile changes also, and becomes positive on the
SE, and negative on the NW. The measured $H_4$ profile becomes
slightly more negative ($\sim -0.1$) at these radii. As discussed in
Sec.~\ref{Paschen}, the presence of young stars in the polar disk may
partially account for the measured increase in the velocity
dispersion profile along the minor axis. Once we consider these
systematic effects on $\sigma$ due to an increase of the Paschen EWs
(Sec.\ref{Paschen}), the minor axis measurements are consistent with a
flat or slightly rising velocity dispersion profile for $r > 4''$
(Fig.\ref{RCmin}). This interpretation may still not be the unique:
the coherent variation observed in $v$, $\sigma$, $H_3$ and $H_4$
profiles for $r>4''$ may also be related to the spheroid structure.

\section{Discussions and conclusions}\label{conclu}

We have obtained new high angular resolution, high signal-to-noise
spectra along the photometric axes of the stellar spheroid in
NGC~4650A. Compared to previous kinematics studies, the new
observations show {\it i)} a kinematic signature of a decoupled core,
and {\it ii)} a non zero rotational velocity and an increasing
velocity dispersion along the minor axis.  In the following we discuss
them and address the main conclusions about the structure and
formation of NGC~4650A.

\subsection{Comparison with previous observations and dynamical models}

We now discuss the current rotation velocities and velocity dispersion
measurements with those previously published in the literature
(Whitmore et al. 1987, Sackett et al. 1994, Swaters \& Rubin
2003). Our data represent a significant improvement in angular
resolution, which is needed for the understanding of the nuclear
regions, and no velocity dispersion measurements were available for
the spheroid minor axis prior to this work.  The rotation curve along
the polar disk at PA$=158^\circ$ was measured from the H$\alpha$
emission line (Whitmore et al. 1987), and from optical absorption
lines at PA$=155^\circ$ (Swaters \& Rubin 2003), with slits aligned
separately along North and South part of the polar disk, so not
through the center of the spheroid; both were
obtained at a slightly different PA from the one of the minor axis
spheroid (PA$=152^\circ$; Iodice et al. 2002).

\begin{figure*}
\includegraphics[width=17cm]{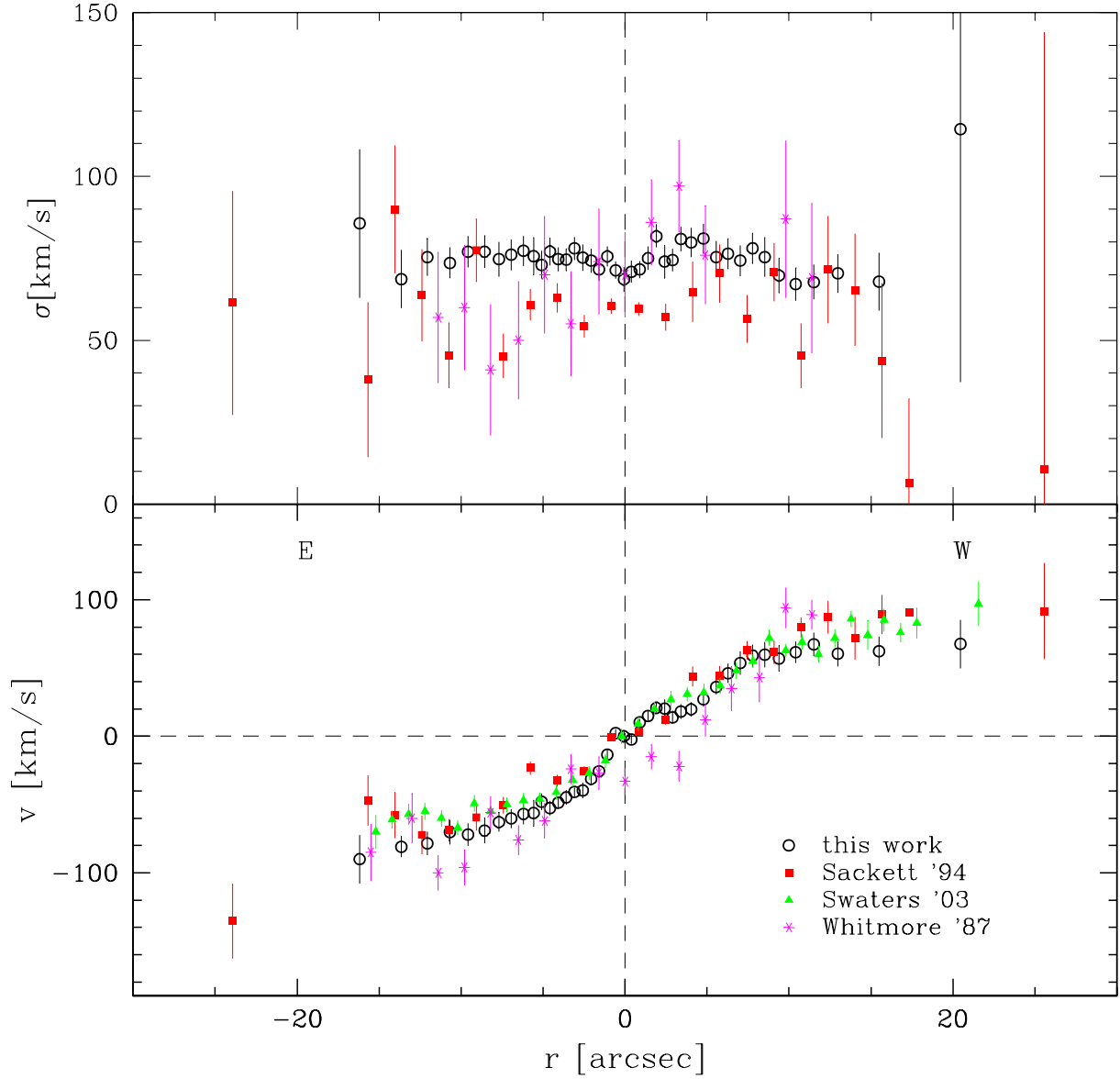}
\caption{Rotation curve and velocity dispersion profile along the
major axis of the stellar spheroid obtained in this work compared with
those in the literature, by Whitmore et al. 1987 ($P.A.= 63^{\circ}$);
Sackett et al. 1994 ($P.A.= 61^{\circ}$); Swaters \& Rubin 2003
($P.A.= 63^{\circ}$).
\label{conf}}
\end{figure*}

A direct comparison of the major axis kinematics data, $v$ and
$\sigma$ is shown in Fig.~\ref{conf}; previous published data show a
larger scatter for the velocity dispersion measurements, while the
agreement is good for the rotation curves. Our kinematic data are
consistent within the errors: the new rotation curve is in good
agreement with the previous ones and our $\sigma$ profile has clearly
benefited from a substantial increase in S/N, thanks to the collecting
area of an 8 meter telescope. The new $\sigma$ measurements are in
agreement with those by Whitmore et al. (1987), while they are
systematically larger than the measurements from Sackett et
al. (1994).  The average difference in this case is about $15$ \kms,
similar to the 19 \kms, which Sackett et al. (1994) subtracted off the
data to account for their systematics.

The new high resolution CaT spectra are better tracers of the
kinematics for the NGC~4650A spheroid than was available before. The
higher angular resolution and the net increase in S/N allow us to 1)
measure a {\it flat} velocity dispersion profile along the spheroid
major axis, while previous $\sigma$ measurements were too scattered to
reliably establish any trend with radius, 2) detect a decrease of the
velocity dispersion in the center, and 3) hint of increasing $\sigma$
in the NE, as previously noticed by Sackett et al. (1994).

The measured flat velocity dispersion profile along the spheroid major
axis shows that both the linear decreasing fit $\sigma_r(r) =
\sigma_r(0) - K\times r_d $ proposed by Sackett et al. (1994) and the
exponential empirical law $ \sigma_r(r) = \sigma_r(0)\exp[-(r/4r_d)^4]$
proposed by Combes \& Arnaboldi (1996) do not reproduce the observed
trend with radius. Previous conclusions that the same authors drew were
based on data that are no longer valid, and the dynamical model for NGC
4650A must be revised.

\subsection{The spheroid kinematics}\label{sphkin}

{\it The stellar spheroid $v/\sigma - \epsilon$ relation} - The light
distribution of the central stellar spheroid in NGC~4650A is fit by a
nearly exponential surface brightness profile, with $r_e = 6''.7$ and
$\epsilon = 0.46$ in the I band, and nearly edge-on as inferred from
the presence of a thin disk aligned with the spheroid's major axis
(Iodice et al. 2002). This is a low luminosity spheroid, with $M_B
\sim -18.1$ (Gallagher et al. 2002) based on HST photometry.

The observed kinematics along the major axis is consistent with
that of a spheroidal galaxy supported by rotation: at $r\geq 2r_e$,
where the ellipticity $\epsilon \sim 0.5$, we estimate $(v/\sigma)^*
\sim 0.96$.  The anisotropy parameter $(v/\sigma)^*$ is defined as the
ratio of the observed value of $(v/\sigma)$ and the theoretical value
for an isotropic oblate rotator
$(v/\sigma)_{oblate}=[\epsilon/(1-\epsilon)^{1/2}]$ (Binney 1978).
For a total B magnitude of $M_B \sim -18.1$ (Gallagher et
al. 2002), in the plane $\log(v/\sigma)^* - M_B$ the central spheroid
of NGC~4650A is located in the region of low luminosity elliptical
galaxies flattened by rotation (see Fig.18 in Bender et al. 1994).

\begin{figure*}
\includegraphics[width=15cm]{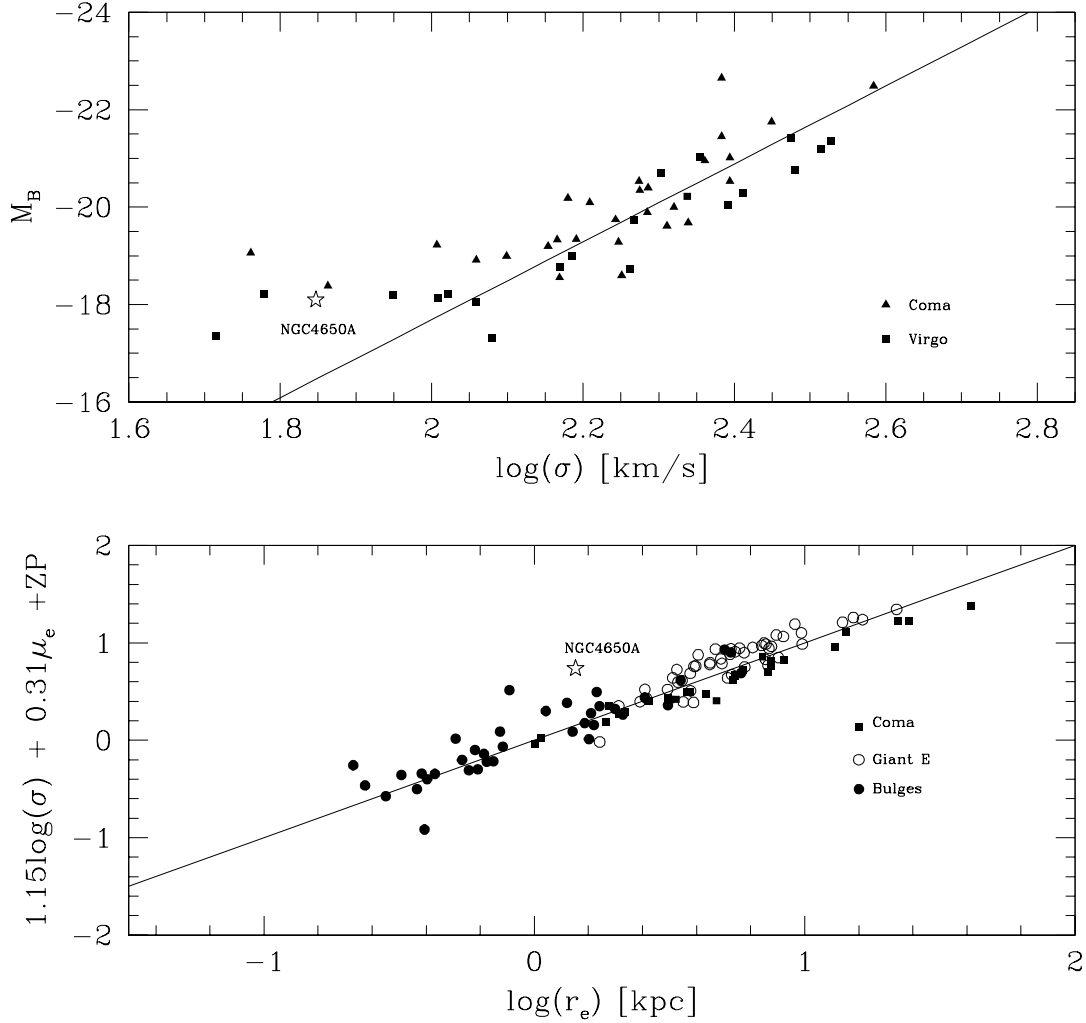}
\caption{Top panel - Faber-Jackson relation between the central
velocity dispersion and the absolute B magnitude. Local early-type
galaxies in the Virgo (squares) and Coma (triangles) clusters are from
Dressler et al. (1987); the straight line is the principal components
fits of these data by Ziegler \& Bender 1997); the position of the
spheroid in NGC~4650A is labeled on the figure.  Bottom panel - The
B-band FP for different morphological types of galaxies: early-type
galaxies in the Coma cluster are from Jorgensen et al. 1996; giant
ellipticals are from Bender et al. 1992; bulges of disk galaxies are
both from Falcon-Barroso et al. 2002 and Bender et al. 1992). The
straight line is the fit performed by Falcon-Barroso et al. 2002 which
include both Coma galaxies and bulges. The position of the stellar
spheroid in NGC~4650A is shown on the figure.
\label{FJFP}}
\end{figure*}

{\it Faber-Jackson relation and the Fundamental plane } - We have
investigated the position of the stellar spheroid in NGC~4650A with
respect to the {\it Faber-Jackson relation (FJR)} (Faber \& Jackson
1976) and {\it Fundamental Plane (FP)} (Djorgovski \& Davies 1987) of
elliptical galaxies. In Fig.\ref{FJFP} (top panel) we compare the
absolute B magnitude and the central velocity dispersion for the
spheroid in NGC~4650A with those for ellipticals in the Virgo and Coma
clusters (in the sample studied by Dressler et al. 1987).  For the
observed velocity dispersion, the spheroidal component in NGC~4650A is
more luminous with respect to the predicted value by the best-fit of
FJR (which did not include the 5 galaxies with $log(\sigma)<2$).

Respect to the B-band\footnote{The effective radius and the effective
surface brightness used to derive the FP for the spheroid in NGC~4650A were
estimated by fitting the whole HST B-band light profiles with a Sersic
law ($I(r)\propto r^{1/n}$), where $n \sim 2$, $\mu_e \sim 23.00 \pm
0.02$ and $r_e \sim 7.1 \pm 0.1$ arcsec.}  FP of elliptical galaxies
and bulges (by Bender et al. 1992; Jorgensen et al. 1996;
Falcon-Barroso et al. 2002), the spheroid in NGC~4650A falls slightly
above the average relation for Es and S0 in the Coma cluster (see
bottom panel of Fig.\ref{FJFP}). Forbes \& Ponman (1999) showed that
bluer early-type galaxies have larger deviations from the FJR: the
observed deviations of the NGC~4650A spheroid from both the FJR and
FP, toward higher luminosity, is consistent with the stellar component
being younger and bluer (Gallagher et al. 2002; Iodice et al. 2002)
than the one in standard early-type galaxies.

{\it Kinematics on the spheroid minor axis} - The new kinematics
observed along the minor axis is rather complex and needs to be
discussed in some details.

In the inner regions ($r\le 4''-6''$), where the spheroid's light
dominates, the observed kinematics is consistent with an isotropic
oblate spheroid, as suggested by the major axis kinematics. The LOS
velocities are close to zero and the velocity dispersions measurements
are very similar to the constant values measured along the major axis
slit. Within this regions, the slightly asymmetric circular velocity
and dispersion profiles, with respect to the galaxy center, may
reflect the contamination by the ``inner polar disk arms''
(Fig.\ref{fig2}). On the SE side respect to the galaxy center, along
the minor axis slit, the light includes most of the inner polar disk
arms (Gallagher et al. 2002, Iodice et al. 2002) and both the LOS
velocity and the velocity dispersion at $r \ge 0.5''$ are larger than
those observed on the NW side (Fig.\ref{RCmin}).

At larger radii (for $r \ge 6''$), the dispersion increases and larger
rotational velocities are observed. Three possible interpretations are
currently viable: {\it i)} the observed profiles are tracing the
intrinsic spheroid kinematics; {\it ii)} these effects are produced by
contamination of the stellar motion in the polar disk or {\it iii)} by
stronger Paschen lines in a mixed population (Sec.\ref{Paschen}).

If the stellar spheroid has indeed intrinsic minor axis rotation
(Fig.\ref{RCmin}, with $\mu \sim 0.23$\footnote{According to Binney
(1985), the amount of minor axis rotation is parametrized as $\mu =
v_{min}/\sqrt{(v_{maj}^{2} +v_{min}^{2})}$.} at $r\sim 0.5 r_e$), and
the presence of isophotal twist (Fig.\ref{fig2}), both suggest that
the system may be triaxial (Wagner, Bender \& Moellenhoff 1988; Franx,
Illingworth \& de Zeeuw 1991). If the flat or slightly increasing
velocity dispersion profile at large radii is intrinsic, it may trace
the presence of larger masses along the meridian plane.  This would be
consistent with the presence of a massive polar disk (Arnaboldi et
al. 1997; Iodice et al. 2002) and with the empirical evidences for a
dark halo flattened along this direction from the Tully-Fisher
relation of PRGs (Iodice et al. 2003).

The potential generated by a bar can also induce minor axis rotation
(Athanassoula 1992), but the presence of a bar in the stellar spheroid
of NGC 4650A is not supported by the photometry.

\subsection{The nuclear regions of NGC~4650A}\label{nuc}

The stellar spheroid of NGC~4650A is known to contain a very luminous,
compact nucleus, whose extension is about 20 pc ($\sim 0.1''$) and
$L_I \sim 8 \times 10^7 L_{\odot}$, which is superimposed on a more
extended and rounder structure, of about 60 pc (Gallagher et
al. 2002), see also Sec.~\ref{result}.
The high angular resolution of our spectra allows us to reveal the
kinematic signature of the nucleus: within a $0''.5$ radius along the
major axis, we observe a decoupled component in the circular velocity
profile, associated with a lower velocity dispersion.  Such features
indicate the presence of a small core, kinematically distinct from the
main central spheroid, whose extension ($\sim 100$ pc) is consistent
with the photometry. This is also comparable with the sizes of the
kinematically decoupled cores (KDC) observed in ``normal'' early-type
galaxies, which varies from some hundred kpc (Kropolin \& Zeilinger
2000) to about $30 - 60$ pc, for the very small cores detected in
HST images (Krajnovi\'{c} \& Jaffe 2004). The HST data for KDC in
early-type galaxies show that the isophotes in these regions are
rounder and have a different PA with respect to the outer regions
(Krajnovi\'{c} \& Jaffe 2004): this behavior is very similar to what is
observed in the central $0''.5$ radius in NGC~4650A (Fig.~\ref{fig2}).

The observed KDC kinematics in early-type galaxies has very similar
features to those measured in the nuclear regions of NGC~4650A; on the
whole, the velocity profile is characterized by a central asymmetry,
with a flat or a decreasing velocity dispersion profile at the
correspondent nuclear radii.

An upper limit to the mass of the KDC observed in NGC~4650A is $\sim
10^8 M_{\odot}$: here we have assumed that the circular-orbit speed at
$0.5''$ from the center is less than the squared sum of the velocity
dispersion and the measured circular velocity. The nuclear mass
estimate for NGC~4650A is higher than the masses of the star cluster
nuclei derived for late-type spirals, which varies from $10^5$ to
$10^7 M_{\odot}$ (Matthews \& Gallagher 2002; B$\ddot{o}$ker et
al. 2001): as already suggested by Gallagher et al. (2002), this is
consistent with a more luminous star cluster nucleus in NGC~4650A, for
which we estimate a mass-to-light ratio $M/L \sim 1.5
M_{\odot}/L_{\odot}$.

The moderate age for the star cluster nucleus in NGC~4650A (Gallagher
et al. 2002), based on the $B - I$ color, is consistent with the
absence of Paschen lines in the nuclear regions (Sec.~\ref{Paschen}).

KDC are often related to secondary event in a galaxy evolution
(Bertola \& Corsini 1999). Numerical simulations indicate
that kinematically peculiar nuclei may result from the merging of two
disk galaxies, where the size of the KDC becomes smaller (less than
the effective radius) as the mass ratio of the two merging galaxies is
smaller than one (Balcells \& Gonzalez 1998). KDC are also present in
barred galaxies, and are characterized by a low velocity dispersion in
the center (Emsellem et al. 2001). In this case, they are the result
of a secular evolution, rather than of an interaction, because the
gravitational torques by the bar drives the gas toward the central
regions and this then forms new stars with lower velocity dispersion
(Wozniak et al. 2003), but in NGC 4650A, no evidence for a bar-like
structure was detected (Gallagher et al. 2002; Iodice et al. 2002),
and the ellipticity and PA profiles do not show the typical features
observed for barred galaxies (Wozniak et al. 1995; Erwin \& Sparke
1999).

The formation scenarios for PRGs, either via major merging of two disk
galaxies (Bekki 1998b) or accretion of external material (Bournaud \&
Combes 2003), predict that some gas flows to the galaxy center and
quickly forms a small star cluster. This can easily lead to the
formation of a KDC in these systems.

\subsection{Implications for the formation of NGC~4650A} 

As discussed in Sec.~\ref{sphkin}, the observed kinematics and the
photometry suggest that the central galaxy in NGC~4650A is a spheroid
with a nearly exponential surface brightness profile, supported by
rotation.  In a merging scenario with low relative velocity and low
impact parameter (Bekki 1998b, Bournaud \& Combes 2003), a high mass
ratio is required to form a massive polar disk, as observed in
NGC~4650A: the 'victim' mass should exceed the 'intruder'
mass. According to simulations of galaxy mergers (e.g. Bournaud et
al. 2005) this would convert the intruder into an elliptical-like, not
rotationally supported, stellar system. Differently, in the tidal
accretion scenario, with large relative velocity and large impact
parameter, for a particular orbital configuration and a gas-rich
donor, a polar ring and/or disk may form both around a disk or an
elliptical galaxy (Bournaud \& Combes 2003). External gas can be also
accreted from the cosmic web filaments (Dav\'e et al. 2001; Maccio et
al. 2005): in this formation scenario, there is no limits to the mass
of the accreted material, thus a very massive polar disk may develop
either around a stellar disk or a spheroid. 

The new kinematic data obtained for NGC~4650A suggest that the
accretion scenarios are favored, nevertheless more investigations are
needed to solve the discrepancies between the observations and the
theoretical predictions.

\acknowledgments
The authors are very grateful to the referee, V. Rubin, for comments and
suggestions which let us improve this work.  EI wish to thank E. Cappellaro,
F. Patat and J. Alcal\'{a} for the help during data reduction. EI is
also very grateful to E.M. Corsini, N.R. Napolitano, G. Busarello and
F. La Barbera for many useful discussions and suggestions.  We thank
R. Swaters for making available the stellar absorption line rotation
curves (derived from the MgI region of the spectrum) for NGC~4650A. EI
and MC acknowledge financial support by the Italian Ministry of
Education, University and Research (MIUR) through grant COFIN2004020323.
EI and MA acknowledge financial support from INAF, Project of National
Interest (PI: M.A.)  and the Swiss National Science Foundation under
grant 200020-101766. LSS would like to thank the Max-Planck-Institute
for Astrophysics in Garching for their hospitality, and the US
National Science Foundation for support through grant
AST-00-98419. JSG acknowledge the financial support by the US National
Science Foundation, under grant AST-98-03018, and would also like to
thank Basel Observatory for travel support.

\clearpage

\clearpage

\clearpage

\clearpage

\clearpage

\clearpage

\clearpage

\end{document}